# Controlled normal and inverse magnetoresistance and current-driven magnetization switching in magnetic nanopillars.

M. AlHajDarwish,[1] H. Kurt,[1] S. Urazhdin,[1] A. Fert,[2] R. Loloee, W.P. Pratt Jr.,[1] and J. Bass[1]

*[1] Department of Physics and Astronomy, Center for Sensor Materials, Center for Fundamental Materials Research, Michigan State University, East Lansing, MI, USA 48824-2320.*
*[2] Unite Mixte de Physique, CNRS/THALES, Orsay, France 91404.*

Combining pairs of ferromagnetic metals with different signs of scattering anisotropies, let us independently invert the magnetoresistance and the direction of current-driven switching in ferromagnetic/non-magnetic/ferromagnetic metal nanopillars. We show all four combinations of normal and inverse behaviors, at both room temperature and 4.2K. In all cases studied, the direction of switching is set by the net scattering anisotropy of the fixed (polarizing) ferromagnet. We provide simple arguments for what we see.

The prediction and discovery of current-driven magnetic moment reversal in ferromagnetic/non-magnetic/ferromagnetic (F1/N/F2) nanopillars has stimulated great interest, driven both by fundamental physics questions and by device potential.[1-18] Researchers have varied layer thicknesses, temperature, applied magnetic field H, layer coupling, etc.,[1-18] but have not reversed spin-dependent scattering anisotropies within F1 and/or F2 or at their interfaces with N.

All data published so far are for samples where minority electrons are scattered more strongly than majority, both within F1 and F2 and at their interfaces with N—positive scattering anisotropies. The nanopillar's resistance is then smallest in high magnetic fields, H, where the magnetizations in F1 and F2, **M1** and **M2,** are aligned parallel (P)—normal magnetoresistance (MR), and positive dc current (I > 0) flowing from F1 to F2 causes **M2** to switch from P to anti-parallel (AP) to **M1**—normal switching. Reversing the net scattering anisotropies for both F1 and F2, so majority electrons are scattered most strongly by both, is known to leave the MR normal.[19,20] Here net scattering anisotropy for F1 means the resultant effect on the MR of anisotropy from the bulk of F1 and its F1/N interfaces (or the same for F2 and N/F2). Conversely, at 4.2K, combining net positive scattering anisotropy for F1 with net negative anisotropy for F2 (or vice-versa) has been shown to invert the MR, so that the resistance is largest in the P state.[19,20] But controlled inversion of the direction of switching by manipulating scattering anisotropies has not yet been shown.

In this Letter, we show that combining metals with different scattering anisotropies in F1, F2, and their interfaces with N, lets us invert either just the MR, just the switching, or both together, at 4.2K and room temperature (295K). These inversions confirm that the switching is not due to the self-Oersted field, for which scattering anisotropies are irrelevant. We show that our results can be understood with simple arguments.

Our sample preparation and measuring techniques are described elsewhere.[18] Our multilayers were triode sputtered onto Si substrates, and patterned into nanopillars of approximately elliptical shape and dimensions ~ 70 nm x 130 nm. The samples consisted of a thick Cu lower contact, the multilayer, and a thick Au top contact. The N-layer was made thick (6-20 nm) to minimize exchange coupling between F1 and F2. To simplify switching, the sample was ion-milled only through F2 and part of N, leaving F1 (fixed polarizer) to have much larger area (~ $\mu m^2$) and to be thicker than F2. Dipolar coupling between F1 and F2 is then minimal, and H reverses **M1** and **M2** sequentially**,** but I reverses only **M2** of F2 (free switcher). Alloys were made by sputtering either from alloy targets or from pure metal targets containing impurity plugs. Differential resistances, dV/dI, were measured with four probes and lock-in detection with an ac current ~ 20 μA at 8 kHz. H was along the nanopillar easy axis. For each combination of MR and switching, data at 295K and 4.2K show that the switching direction is independent of temperature. Each switching behavior was also independently reproduced, and no inconsistent switching was seen.

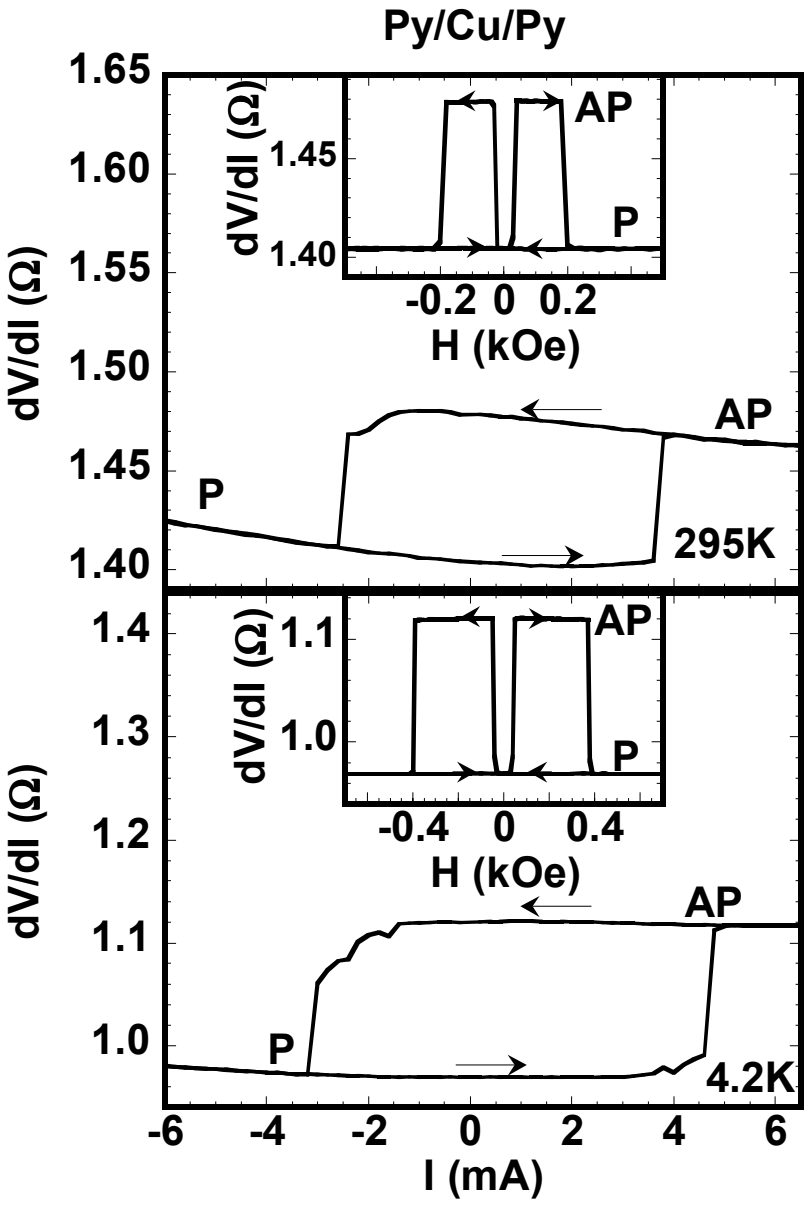

Fig. 1. Py(24)/Cu(10)/Py(6) data at 295K (top) and 4.2K (bottom) showing normal MR (dV/dI vs H at I = 0) in the insets and normal switching for dV/dI vs I in the main figures at H = 0 Oe for 295K and at H = 20 Oe for 4.2K.

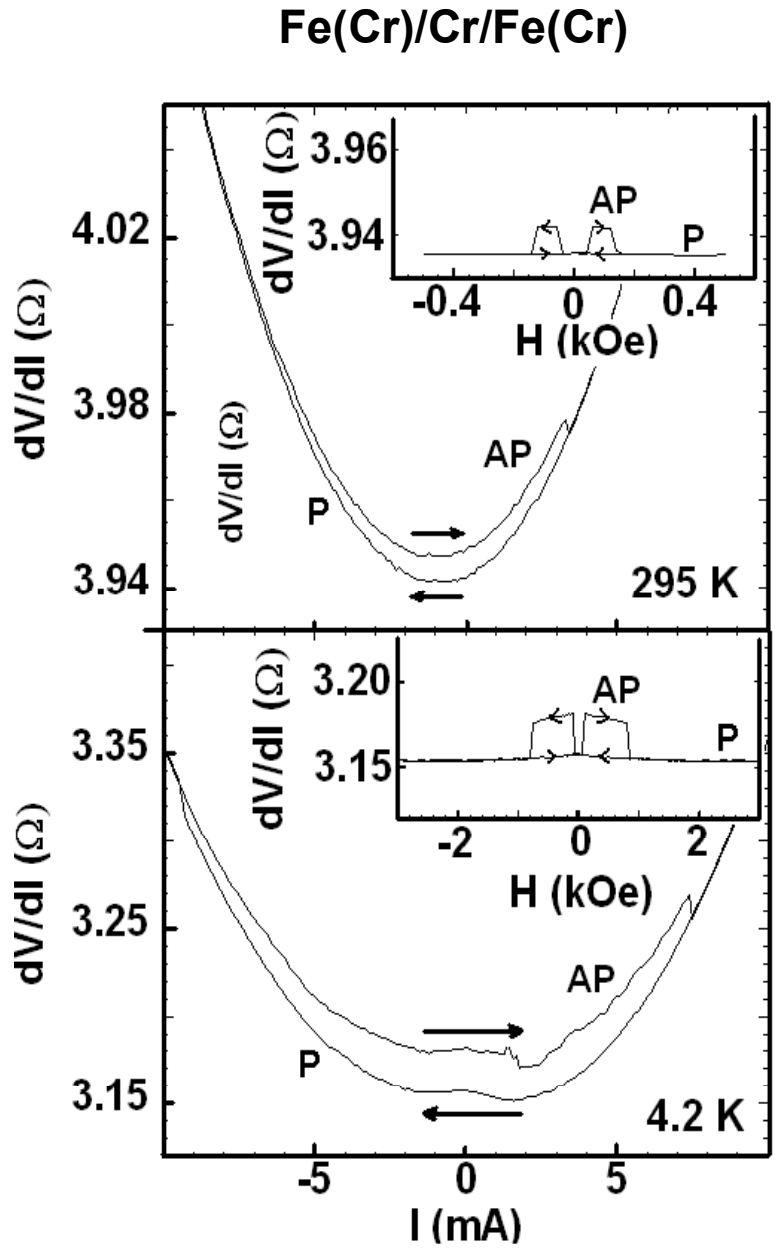

Fig. 2. Fe(Cr)(30)/Cr(6)/Fe(Cr)(3) data at 295K (top) and 4.2K (bottom) showing normal MR (dV/dI vs H at I = 0) in the insets but inverse switching for dV/dI vs I at H = 0 in the main figures.

Permalloy (Py = $Ni_{84}Fe_{16}$) and Py/Cu interfaces both have positive scattering anisotropy.[19] As expected, Fig. 1 shows that Py(24)/Cu(10)/Py(6) nanopillars (layer thicknesses are in nm) give normal MR and normal switching. The data are similar to those for a Py(20)/Cu(10)/Py(6) nanopillar.[18] At both temperatures, the MR transitions from P to AP occur after H passes through zero, consistent with little or no magnetic coupling The agreement between the minimum and maximum values of dV/dI for the MR and current-driven curves also shows that the switching is complete. Figs. 2-5 show similarly weak coupling and complete switching.

In contrast to Py and Py/Cu, Fe(Cr) = $Fe_{95}Cr_5$ and Fe/Cr interfaces both have negative scattering anisotropies.[20-26] Since F1 and F2 are the same alloy, Fe(Cr)(30)/Cr(6)/Fe(Cr)(3.5) nanopillars should give normal MR.[19,20] Fig. 2 shows that they do, and also give inverse switching. The changes in dV/dI vs I or H are smaller than for Py/Cu/Py, due to spin-memory-loss in the Cr(6) layer[22] and smaller scattering anisotropy of Fe(Cr).[20] More Fe(Cr)/Cr/Fe(Cr) data will be given in [21].

Fig. 3 shows that the four component system Py(20)/Cu(7)/Cr(3)/Fe(Cr)(3) gives inverse MR with normal switching. Combining the net positive scattering anisotropy for F1 with the net negative anisotropy for F2 gives the expected inverse MR.[19,20] But the switching is normal—I > 0 switches from P to AP—remembering that inverse MR means largest resistance in the P state.

Fig. 4 shows the fourth alternative, with Ni(Cr)(20)/Cu(20)/Py(10), where 20 nm of Ni(Cr) = $Ni_{97}Cr_3$ is thick enough so its negative anisotropy dominates the MR over the positive anisotropy of the Ni(Cr)/Cu interface.[27] Combining the net negative anisotropy for Ni(Cr) with the positive anisotropies for Py and Py/Cu, gives the expected inverse MR, and now inverse switching.

Fig. 5 shows another way to achieve inverse MR with normal switching, with Py(24)/Cu(10)/NiCr(4). In both Figs. 3 and 5, this combination occurs with negative net anisotropy for F2. However, the interface anisotropies of N/F2 are opposite—negative in Fig. 3 but positive in Fig. 5.

Before turning to theory, we summarize the results in Figs. 1-5. As expected for the MR,[19,20] when the net scattering anisotropies for F1 and F2 are the same (Figs. 1, 2), the MR is normal, and (shown for the first time at 295K) when they are opposite (Figs. 3-5), the MR is inverse. Shown also for the first time, when the net scattering anisotropy for F1 is positive, switching for I > 0 is normal (Figs. 1,3,5), and when it is negative (Figs. 2,4), switching for I > 0 is inverse. The direction of switching is independent of the net scattering anisotropy of F2. Figs. 3 and 5 show that the switching is also independent of the scattering anisotropy of N/F2. Finally, dominance by the bulk contribution of scattering anisotropy in Ni(Cr) is inconsistent with ballistic transport through the nanopillar, since in ballistic transport the interfaces must dominate the scattering. Thus, at least for samples containing Ni(Cr), diffusive scattering must be included.

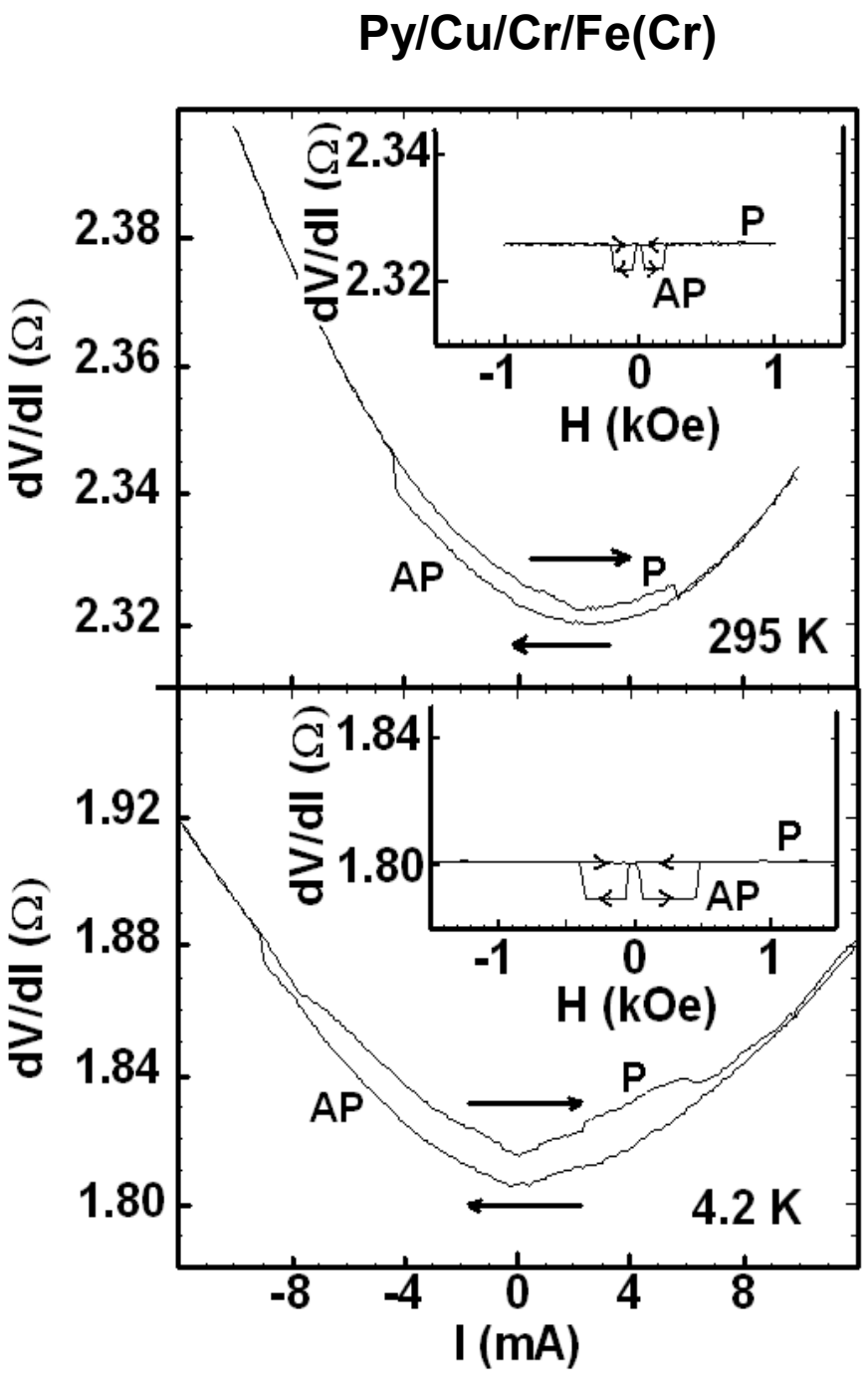

Fig. 3. Py(20)/Cu(7)/Cr(3)/Fe(Cr)(3) data at 295K (top) and 4.2K (bottom) showing inverse MR (dV/dI vs H at I = 0) in the insets but normal switching for dV/dI vs I at H = 0 in the main figures.

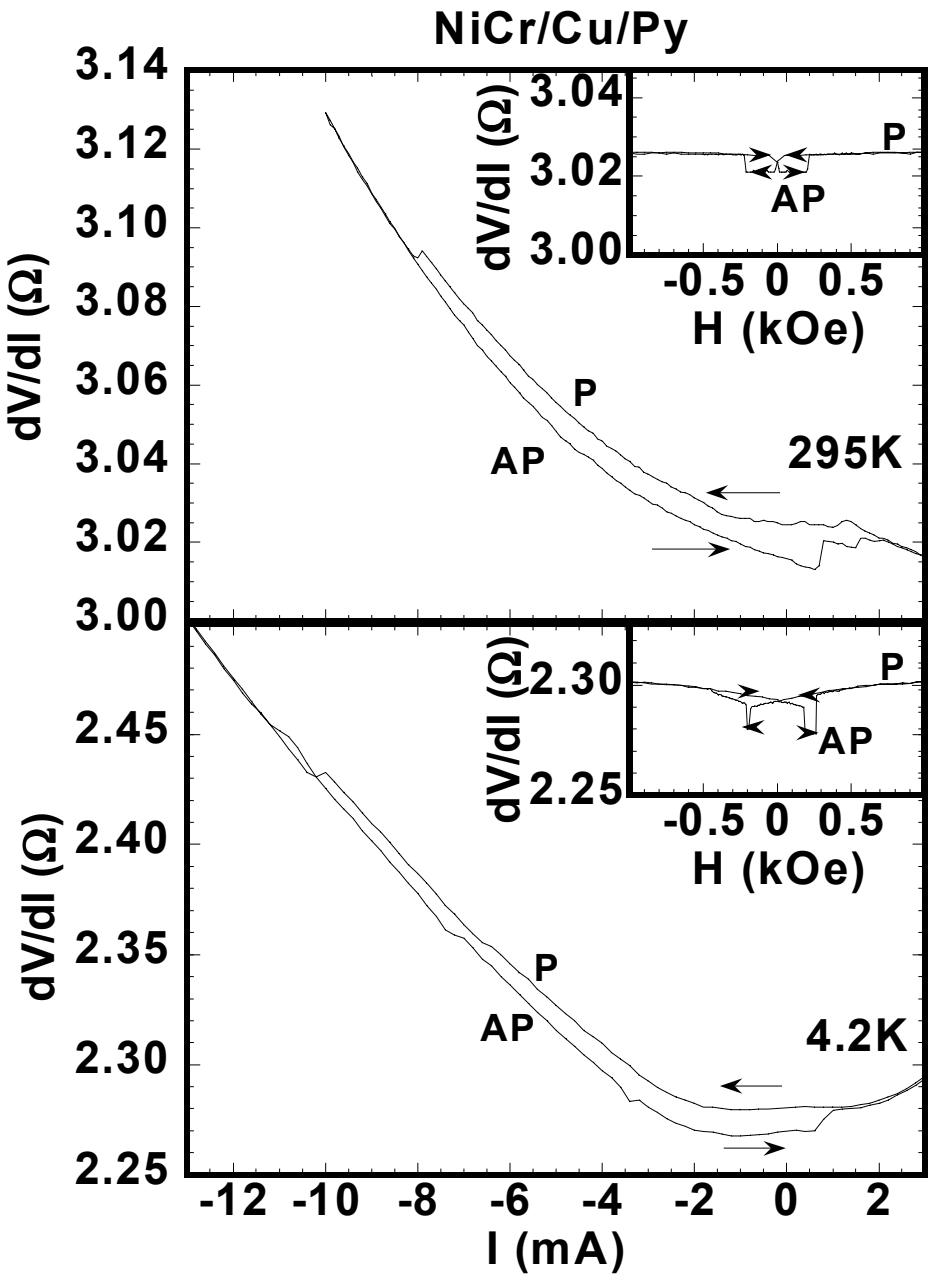

Fig. 4. Ni(Cr)(20)/Cu(20)/Py(10) data at 295K (top) and 4.2K (bottom) showing inverse MR (dV/dI vs H at I = 0) in the insets and inverse switching for dV/dI vs I at H = 0 in the main figures.

Intriguingly, the switching directions in Figs. 1-5 accord with the simplest ballistic spin-transfer picture,[1,16] where F1 just polarizes I (positively for positive net scattering anisotropy and negatively for negative net scattering anisotropy), and then positively polarized I > 0, (or negatively polarized I < 0) reverses **M2** from P to AP to **M1**. Waintal et al.[10] argue that the switching direction being independent of the properties of F2 is more general. However, as these models assume ballistic transport throughout and, at least in Ni(Cr) the transport must be diffusive, they cannot fully describe our data.

For diffusive transport, the polarization in N depends upon the scattering anisotropies of both F1 and F2, and the switching depends upon both spin-polarized charge current and spin-accumulation effects.[13,17] We input the best available parameters[19,22,27] into Eq. 5 of ref. [17], where the torque is given by a weighted sum of contributions from both spin accumulation and spin current terms, evaluated both within N at the F2/N interface and within F1 at the N/F1 interface. In all cases except Fig. 4, the signs of both the polarized current and spin-accumulation contributions are the same as that set by the net spin anisotropy of F1, just as in the simplest picture. For Fig. 4, the situation is more complex, because Py as F2 reverses the current polarization in the P state. For P to AP switching, the sign of that polarization is positive, opposite to that for F1 alone, but the spin accumulation term brings back our observed result. For AP to P switching, the spin-polarization is negative and dominates a small spin accumulation term. Thus, we reproduce the behaviors in Fig. 4.

To summarize, we have shown that judiciously chosen pairs of ferromagnetic metals produce all four combinations of normal and inverse MR and current-driven switching, at both 4.2K and 295K. For the samples studied, the switching direction is determined solely by the net anisotropy for F1. These results, which are characteristic of spin-anisotropy effects, rule out self-Oersted field switching. Intriguingly, our observed switching directions can be understood either by simple arguments involving ballistic transport, or by diffusive transport. In the latter case, spin-accumulation plays an important role for the P to AP transition in Fig. 4, where the dominance of the bulk anisotropy of NiCr in the MR also clearly requires a diffusive treatment of scattering. For devices, the standard normal-normal system gives the largest MR and the largest change in dV/dI upon switching.

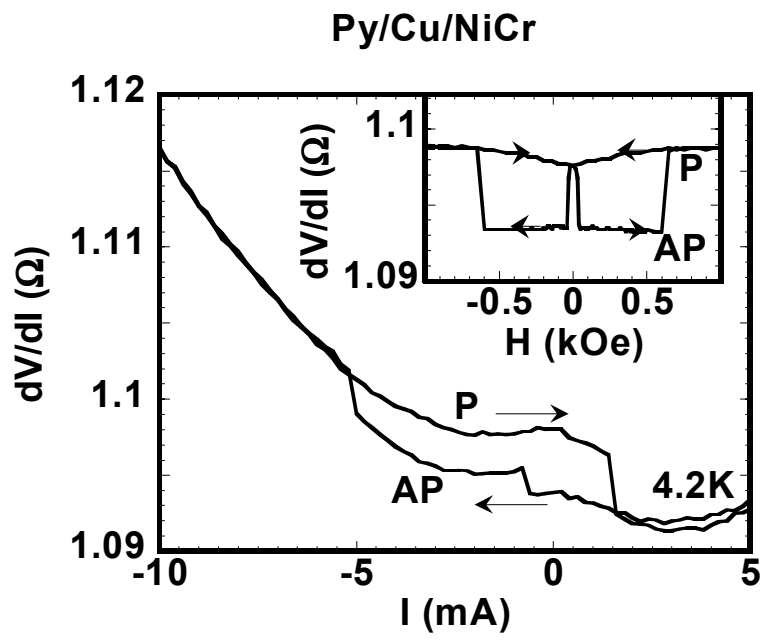

Fig. 5. Py(24)/Cu(10)/NiCr(4) data at 4.2K showing inverse MR (dV/dI vs H at I = 0) in the inset and normal switching for dV/dI vs I at H = 0 Oe in the main figure.

Acknowledgments: The authors thank Henri Jaffrès for spin-accumulation calculations, and N.O. Birge, P.M. Levy and M.D. Stiles for helpful suggestions. This research was supported by the MSU CFMR, CSM, Keck Microfabrication Facility, NSF grants DMR 02-02476, 98-09688, NSF-EU collaborative grant 00-98803 and Seagate Technology.

--------------------